\documentclass[letterpaper,english,aps,prl,letterpaper,twocolumn,showpacs,floatfix,groupedaddress,superscriptaddress]{revtex4}
\usepackage{mathptmx}
\usepackage[T1]{fontenc}
\usepackage[latin9]{inputenc}
\usepackage{amsmath}
\usepackage{amssymb}
\usepackage{mathrsfs}
\usepackage{graphicx}
\usepackage{subfigure}

\makeatletter

\newcommand{\noun}[1]{\textsc{#1}}

\makeatother

\usepackage{babel}

\begin{document}

\title{Relaxation and dephasing in open quantum systems time-dependent density functional theory: Properties of exact functionals from an exactly-solvable model system.}

\author{David G. Tempel}

\address{Department of Physics, Harvard University,
17 Oxford Street, 02138, Cambridge, MA}

\author{Al\'an Aspuru-Guzik}

\address{Department of Chemistry and Chemical Biology, Harvard University,
12 Oxford Street, 02138, Cambridge, MA}

\email{aspuru@chemistry.harvard.edu}

\begin{abstract}
The dissipative dynamics of many-electron systems interacting with a thermal environment has remained a long-standing challenge within time-dependent density functional theory (TDDFT). Recently, the formal foundations of open quantum systems time-dependent density functional theory (OQS-TDDFT) within the master equation approach were established. It was proven that the exact time-dependent density of a many-electron open quantum system evolving under a master equation can be reproduced with a closed (unitarily evolving) and non-interacting Kohn-Sham system. This potentially offers a great advantage over previous approaches to OQS-TDDFT, since with suitable functionals one could obtain the dissipative open-systems dynamics by simply propagating a set of Kohn-Sham orbitals as in usual TDDFT. However, the properties and exact conditions of such open-systems functionals are largely unknown. In the present article, we examine a simple and exactly-solvable model open quantum system: one electron in a harmonic well evolving under the Lindblad master equation. We examine two different representitive limits of the Lindblad equation (relaxation and pure dephasing) and are able to deduce a number of properties of the exact OQS-TDDFT functional. Challenges associated with developing approximate functionals for many-electron open quantum systems are also discussed.

\end{abstract}


\maketitle

\section{I. Introduction}

Due to its attractive balance between accuracy and efficiency, time-dependent density functional theory (TDDFT) is an ideal method for computing the real-time dynamics of many-electron systems~\cite{octopus 1, octopus 2, octopus 3, Castro propagators, Baer nanolett, baer energy loss, baer linear response, van voorhis, annu, review of burke, Rubio transport, Gross 08, kurth 08, Thiele 2008, L.K. 95, strong field}. In its original formulation by Runge and Gross~\cite{runge gross}, TDDFT addresses the isolated dynamics of electronic systems evolving unitarily under the time-dependent Schr{\"o}dinger equation. However, there exist many situations in which the electronic degrees of freedom are not isolated, but must be treated as a subsystem imbedded in a much larger thermal bath. Several important examples include vibrational relaxation of molecules in liquids and solid matrices~\cite{Mukamel Non-Markov,Mukamel Frank-Condon,  Nitzan}, photo-absorption of chromophores in a protein bath~\cite{Redfield Absorption, Redfield Absorption 2, protien dynamics, mukamel 1}, nonlinear spectroscopy in the condensed phase~\cite{scholes, Mukamel book}, electron-phonon coupling in single-molecule transport~\cite{burke, gebauer chemphyschem, gebauer prl, chen 1, chen 2} and exciton and energy transfer~\cite{exciton transfer, nano letter, ENAQT, mukamel 2}. 

In the above examples, the theory of open quantum systems (OQS) within the master equation approach is often used to model the dissipative electron dynamics~\cite{nakajima, zwanzig, breuer, Kossakowski72a, lindblad, gorini}. In the master equation approach, one traces over the bath degrees of freedom, arriving at a simpler description in terms of the reduced density matrix of the electrons only. The price paid is that the resulting dynamics are non-unitary, and in general the interaction of the system with it's environment must be modeled in an approximate way.  Even with simple system-bath models, the exact solution of the master equation for the reduced dynamics of an interacting many-electron system is computationally intractable. Therefore, open quantum systems TDDFT (OQS-TDDFT) offers an attractive approach to the many-body open-systems problem. 

Several different formulations of OQS-TDDFT have been proposed in the last few years~\cite{burke, tempel, chapter, d prl, d prb, da prb, joel's paper, prl}. In~\cite{burke}, an OQS Runge-Gross theorem was established for Markovian master equations of the Lindblad form. A scheme in which the many-body master equation is mapped onto a non-interacting Kohn-Sham master equation was proposed for application to single-molecule transport. A Kohn-Sham master equation was also used in formulating the linear-response version of OQS-TDDFT, giving access to linewidths of environmentally broadened spectra~\cite{tempel}. In~\cite{tempel}, OQS Casida-type equations were derived and used to calculate the spectrum of an atom interacting with a photon bath. A different formulation of OQS-TDDFT based on the stochastic Schrodinger equation rather than the master equation has also been developed~\cite{d prl, da prb, d prb}. 

Recently, in ~\cite{joel's paper, prl}, the OQS Runge-Gross theorem was extended to arbitrary non-Markovian master equations and a Van Leeuwen construction was established, thereby proving the existence of an OQS-TDDFT Kohn-Sham scheme~\cite{van leeuwen}. We showed that the time-dependent density of an interacting OQS can be reproduced with a non-interacting and closed (unitarily evolving)  Kohn-Sham system. In principle, the closed Kohn-Sham scheme is remarkably useful for real-time dynamics, since it allows one to calculate any property of a many-body OQS by unitarily propagating a set of one-particle orbital equations evolving in a local potential. With suitable functionals, such a scheme could readily be implemented in existing real-time TDDFT codes~\cite{octopus 1, octopus 2, octopus 3}. In practice, the OQS-TDDFT exchange-correlation potential is very complicated object. Not only does it have initial-state and memory dependence as in usual TDDFT~\cite{neepa history, neepa memory, neepa memory perturbations, baer memory}, but it must also be a functional of bath quantities such as the bath spectral density.

In the present manuscript, we study exact features of the OQS-TDDFT closed Kohn-Sham scheme using an exactly-solvable one-electron model system. By focusing on a single electron, we are able to isolate the part of the exact functional arising solely from interaction with the bath, without the need to describe electron-electron interaction effects within the system.

The paper is organized as follows. In Section II, we review the theory of OQS for a many-electron system and the closed Kohn-Sham scheme presented in previous work~\cite{joel's paper, prl}.  Section III presents the model system to be analyzed and discusses the procedure used to obtain exact Kohn-Sham quantities. Section IV presents results and an analysis. The paper concludes with an outlook and discussion of challenges for OQS-TDDFT in Section V. Atomic units in which $e = \hbar = m_{e} = 1$ are used throughout.

\section{II. Open Quantum Systems TDDFT using unitary propagation}

In this section, we begin by briefly reviewing the master equation approach for many-body OQS. We first discuss the most general master equation including non-Markovian effects and initial correlations~\cite{tannor}. We then introduce the Markov approximation and the widely used Lindblad master equation~\cite{breuer, lindblad, Kossakowski72a, van Kampen, redfield original, redfield original 2}. Lastly, we review the construction discussed in~\cite{prl}, where the time-evolving density of an interacting OQS is reproduced with a noninteracting and closed Kohn-Sham system.

\subsection{A. The formally exact many-body master equation}

The starting point of many-body OQS is the unitary evolution for the full density matrix of the system and the reservoir (we use the terms "reservoir" and "bath" interchangeably throughout) ,

\begin{equation}
\frac{d}{dt} \hat{\rho}(t) = -\imath [\hat{H}(t), \hat{\rho}(t)]. 
\label{full dynamics}
\end{equation}

The full Hamiltonian is given by

\begin{equation}
\hat{H}(t) = \hat{H}_{S}(t) + \hat{H}_{R} + \hat{V}.
\label{full hamiltonian}
\end{equation} 

Here,

\begin{equation}
\hat{H}_{S}(t) = -\frac{1}{2} \sum_{i=1}^N \nabla_{i}^2 + \sum_{i<j}^{N} \frac{1}{|\mathbf{r}_i - \mathbf{r}_j|} + \sum_i v_{ext}(\mathbf{r}_i, t),
\label{many-body hamiltonian}
\end{equation}

is the Hamiltonian of the electronic system of interest in an external potential $v_{ext}(\mathbf{r}, t)$. This potential generally consists of a static external potential due to the nuclei and an external time-dependent driving field.  The system-bath coupling, $\hat{V}$, is generally assumed to be weak and is treated using perturbation theory in most applications. $\hat{V}$ acts in the combined Hilbert space of the system and reservoir and so it couples the two subsystems. $\hat{H}_{R}$ is the Hamiltonian of the reservoir, which typically has a dense spectrum of eigenstates relative to the system. The density of states of $\hat{H}_{R}$ determines the decay rate of reservoir correlation functions, whose time-scale in turn determines the reduced system dynamics.

Defining the reduced density operator for the electronic system alone by tracing over the reservoir degrees of freedom, 

\begin{equation}
\hat{\rho}_S(t) = Tr_R\{ \hat{\rho}(t)  \},
\end{equation}

one arrives at the formally-exact quantum master equation,

\begin{equation}
\frac{d}{dt} \hat{\rho}_{S}(t) = -\imath [\hat{H}_{S}(t), \hat{\rho}_{S}(t)] + \int_{t_{0}}^{t} d\tau  \breve{\Xi}(t-\tau) \hat{\rho}_{S}(\tau) + \Psi(t).
\label{non-markovian}
\end{equation}

Here, $\breve{\Xi}(t-\tau)$ is the memory kernel and $\Psi(t)$ arises from initial correlations between the system and its environment. It is referred to as the inhomogeneous term. The above equation is still formally exact, as $ \hat{\rho}_{S}(t)$ gives the exact expectation value of any observable depending only on the electronic degrees of freedom. In practice, however, approximations to $\breve{\Xi}$ and $\Psi$ are required. Of particular importance in TDDFT is the time-dependent electronic density,

\begin{equation}
n(\mathbf{r}, t) = Tr_S \{ \hat{\rho}_{S}(t) \hat{n}(\mathbf{r}) \},
\label{density}
\end{equation}

where $\hat{n}(\mathbf{r}) = \sum_i^{N} \delta(\mathbf{r} - \hat{\mathbf{r}}_i)$ is the number density operator for the electronic system. For OQS, the continuity equation is not strictly satisfied and is modified to

\begin{eqnarray}
&& \frac{\partial}{\partial t} n(\mathbf{r}, t) =-\mathbf{\nabla}\cdot Tr_{S}\{ \hat{\rho}_{S} (t)\mathbf{\hat{j}} (\mathbf{r}) \} \nonumber \\ &+& Tr\{\hat{n}(\mathbf{r})\left( \int_{t_{0}}^{t} d\tau  \breve{\Xi}(t-\tau) \hat{\rho}_{S}(\tau)+ \Psi(t) \right)\}.
\label{pseudocontinuity}
\end{eqnarray}

Here, the first term is the divergence of the usual current arising in closed systems, and is referred to as the "Hamiltonian current". The second term is a contribution to the current due to scattering of electrons with particles in the bath and arises from the non-unitary part of the evolution. One may define a "Dissipative current" by~\cite{gebauer prl, diosi}, 

\begin{equation}
-\mathbf{\nabla}\cdot \mathbf{j}_{disp}(\mathbf{r}, t) = Tr\{\hat{n}(\mathbf{r})\left( \int_{t_{0}}^{t} d\tau  \breve{\Xi}(t-\tau) \hat{\rho}_{S}(\tau)+ \Psi(t) \right)\}.
\end{equation}

It will be seen later that the functional dependence of the Kohn-Sham potential on $\mathbf{j}_{disp}(\mathbf{r}, t)$ plays an important role in OQS-TDDFT.

\subsection{B. The Markov approximation and Lindblad master equation}

Without suitable approximations to the memory kernel $\breve{\Xi}(t)$, solving eq.~\ref{non-markovian} is not easier than solving the full system-bath dynamics described in eq.~\ref{full dynamics}. One often invokes the Markov approximation, in which the memory kernel is local in time, i.e. 

\begin{equation}
 \int_{t_{0}}^{t} d\tau  \breve{\Xi}(t-\tau) \hat{\rho}_{S}(\tau) = \breve{D} \hat{\rho}_S(t).
\end{equation}

The Markov approximation is valid when $\tau_S \gg \tau_B$ is satisfied, where $\tau_S$ is the time-scale for the system to relax to thermal equilibrium and $\tau_B$ is the longest correlation time of the bath. Roughly speaking, the memory of the bath can be neglected when describing the reduced system dynamics, because the bath decorrelates from itself before the system has had a chance to evolve appreciably. $\tau_S$ is inversely related to the magnitude of system-bath coupling, and so a weak interaction between the electrons and the environment is implicit in this condition as well. 

The Lindblad form of the Markovian master equation,

\begin{equation}
\breve{D} \hat{\rho}_S(t) =  \sum_{mn} \Bigg\{ L_{mn}  \hat{\rho}_S(t)  L_{mn}^{\dag} - \frac{1}{2} L_{mn}^{\dag} L_{mn} \hat{\rho}_S(t)- \frac{1}{2} \hat{\rho}_S(t) L_{mn}^{\dag} L_{mn} \Bigg\},
\label{Lindblad}
\end{equation}

is constructed to guarantee complete positivity of the density matrix~\cite{breuer}. This is desirable, since the populations of any physically sensible density matrix should remain positive during the evolution.

As written in eq.~\ref{Lindblad}, the Lindblad equation is simply a mathematical construction which is guaranteed to give a positive density matrix. However, the bath operators $L_{nm}$ can be derived microscopically starting from a system-bath Hamiltonian of the form given in eq. ~\ref{full hamiltonian}~\cite{linden}. In general, the operators $L_{nm}$ will describe "jumps" between eigenstates of $\hat{H}_S$ induced by scattering of electrons with bath particles. We will discuss specific forms of the $L_{nm}$ in subsequent sections.

\subsection{C. Open-interacting to closed-noninteracting mapping}

In this section, we briefly outline the proof given in~\cite{prl}, whereby the density of an interacting and open electronic system is reproduced using a closed and non-interacting Kohn-Sham system.

Starting from the many-body master equation given in eq.~\ref{non-markovian}, one considers an auxiliary "primed" system described by a density matrix $\hat{\rho}_{S}'(t)$, evolving with a different Hamiltonian

\begin{equation}
\hat{H}_{S}'(t) = -\frac{1}{2} \sum_{i=1}^N \nabla_{i}^2 + \sum_{i<j}^{N} \frac{\alpha}{|\mathbf{r}_i - \mathbf{r}_j|} + \sum_i v'(\alpha, \mathbf{r}_i, t),
\label{many-body hamiltonian}
\end{equation}

at interaction strength $\alpha$, and with different memory kernel $\breve{\Xi}'(t)$ and initial correlations $\Psi'(t)$. This auxiliary system evolves under the master equation

\begin{equation}
\frac{d}{dt} \hat{\rho}_{S}'(t) = -\imath [\hat{H}_{S}'(t), \hat{\rho}_{S}'(t)] + \int_{t_{0}}^{t} d\tau  \breve{\Xi}'(t-\tau) \hat{\rho}_{S}'(\tau) + \Psi'(t).
\label{non-markovian2}
\end{equation}

Following a construction similar in spirit to~\cite{van leeuwen}, one is able to prove that there \textit{exists} a \textit{unique} local potential $v'(\alpha, \mathbf{r}, t)$ for the auxiliary system, at any interaction strength $\alpha$, and for arbitrary $\breve{\Xi}'(t,t')$ and $\Psi'(t)$ (with some restrictions discussed in~\cite{prl}), such that 

\begin{equation}
n(\mathbf{r}, t) = Tr_S \{ \hat{\rho}_{S}(t) \hat{n}(\mathbf{r}) \} = Tr_S \{ \hat{\rho}_{S}'(t) \hat{n}(\mathbf{r}) \}
\end{equation}

is satisfied for all times. This means that for a fixed open and interacting system, one can always construct an auxiliary system with different electron-electron and system-bath interactions, such that the potential $v'(\alpha, \mathbf{r}, t)$ enforces the correct density evolution. If one sets $\alpha = 0$, but keeps $\breve{\Xi}'(t) = \breve{\Xi}(t)$ and $\Psi'(t) = \Psi(t)$ then the auxiliary system is a non-interacting, but open Kohn-Sham system. This is similar to the construction used in~\cite{burke, tempel}, but encompasses the non-Markovian case as well. However, one may also choose $\alpha = 0$ and $\breve{\Xi}'(t) = \Psi'(t) = 0$, whereby the density of the original open system is reproduced with a closed and non-interacting Kohn-Sham system. In this case, the density matrix in the auxiliary system is a pure state given by

\begin{equation}
\hat{\rho}_S'(t) = \Phi^*(t) \Phi(t),
\end{equation}

where $\Phi(t)$ is a single Slater determinant. This determinant is constructed by propagating a set of single-particle orbital equations,

\begin{equation}
\imath \frac{\partial}{\partial t} \phi_i(\mathbf{r}, t) \Bigg\{ -\frac{1}{2} \nabla^2 + v_{ks}(\mathbf{r}, t )\Bigg\} \phi_i(\mathbf{r}, t)
\end{equation}

as in usual TDDFT. The density of the original interacting and open system is then simply obtained by square-summing the orbitals. i.e.

\begin{equation}
n(\mathbf{r}, t) = \sum_i | \phi_i(\mathbf{r}, t)|^2 = Tr_S \{ \hat{\rho}_{S}(t) \hat{n}(\mathbf{r}) \}.
\end{equation}

In analogy to usual TDDFT, the Kohn-Sham potential is partitioned as 

\begin{equation}
v_{ks}(\mathbf{r}, t) = v_{ext}(\mathbf{r}, t) + v_{h}(\mathbf{r}, t) + v_{xc}^{open}(\mathbf{r}, t), 
\end{equation}

where $v_{h}(\mathbf{r}, t)$ is the Hartree potential and the unknown functional $v^{open}_{xc}(\mathbf{r}, t)$ accounts for electron-electron interaction within the system as well as interaction between the system and bath. Ideally, these two contributions well be additive, so that one may use standard adiabatic functionals to account for electron-electron interaction within the system and construct dissipative bath functionals to account for system-bath coupling. In general,

\begin{equation}
v_{xc}^{open}(\mathbf{r}, t) = v_{xc}^{open}(\mathbf{r}, t)[n, \breve{\Xi}, \Psi, \hat{\rho}_S(0), \Phi(0)].
\label{functional dependence}
\end{equation}

Formally, the open-systems exchange-correlation potential is a functional not only of the density, but also of the memory kernel, inhomogeneous term and initial state of the interacting open system, as well as the initial state of the closed Kohn-Sham system.

\section{III. An exactly-solvable model system}

In this section, we construct the exact OQS-TDDFT Kohn-Sham potential for an exactly-solvable model system: one electron in a harmonic well evolving under the Lindblad equation. Our analysis focuses on two limiting cases of the Lindblad master equation. The first limit is that of pure dephasing without relaxation in which the bath decoheres the system, but no energy is exchanged. The second limit is that of relaxation with no pure dephasing.

\subsection{A. Construction of the exact OQS functional}

With the system Hamiltonian given by

\begin{equation}
\hat{H}_S = -\frac{1}{2} \frac{d^2}{dx^2} + \frac{1}{2} {\omega}^2 x^2,
\label{Hamiltonian}
\end{equation}

the Lindblad equation

\begin{eqnarray}
 \frac{d}{dt} \hat{\rho}_{S}(t) &=& -\imath [\hat{H}_{S}, \hat{\rho_{S}}(t)] \nonumber \\ &+&  \sum_{mn} \Bigg\{ L_{mn}  \hat{\rho}_S(t)  L_{mn}^{\dag} \nonumber \\ & -& \frac{1}{2} L_{mn}^{\dag} L_{mn} \hat{\rho}_S(t)- \frac{1}{2} \hat{\rho}_S(t) L_{mn}^{\dag} L_{mn} \Bigg\}
 \label{HO lindblad}
\end{eqnarray}

can be solved exactly to obtain $\hat{\rho}_{S}(t)$. The exact time-dependent OQS density can then be constructed using eq.~\ref{density}. With an exact OQS density, it is a simple exercise to construct the closed Kohn-Sham system which reproduces this density using a unitary evolution. In this section we work in one dimension, but the formulas apply to higher dimensions as well.

For one electron, there is a single occupied Kohn-Sham orbital given by~\cite{Hessler, correlation, d amico}

\begin{equation}
\phi(x, t) = \sqrt{n(x, t)} e^{\imath \alpha(x, t)}.
\label{orbital}
\end{equation}

By construction, this orbital must evolve under the time-dependent Kohn-Sham equation

\begin{equation}
\imath \frac{\partial}{\partial t} \phi(x, t) = \big\{-\frac{1}{2}\frac{d^2}{dx^2} +v_s(x,t) \big\} \phi(x, t),
\label{ks equation}
\end{equation}

in such a way that the true OQS density is reproduced for all times, i.e.

\begin{equation}
n(x,t) = |\phi(x, t)|^2 = Tr_S \{ \hat{\rho}_{S}(t) \hat{n}(\mathbf{x}) \}.
\end{equation}

Here,

\begin{equation}
v_{ks}(x,t) = v_{ext}(x)+ v_{h}(x, t) + v_{xc}^{open}(x, t)
\end{equation}

is the OQS-TDDFT Kohn-Sham potential. We now substitute eq.~\ref{orbital} in eq.~\ref{ks equation} and use the fact that for one electron, the exchange potential exactly cancels the self interaction in the Hartree potential. The result is an exact expression for the OQS-TDDFT correlation potential in terms of known quantities,

\begin{eqnarray}
 v_c^{open}(x, t) &=& - \frac{\partial}{\partial t} \alpha(x, t) - \frac{1}{2} \left[ \frac{\partial}{\partial x} \alpha(x, t)\right] ^2 \nonumber \\&+& \frac{1}{4 n(x, t)}\frac{\partial^2}{\partial x^2} n(x, t) \nonumber \\ &-& \frac{1}{8 n(x, t)^2}\left[ \frac{\partial}{\partial x} n(x, t)\right]^2 - v_{ext}(x).
 \label{exact correlation}
\end{eqnarray}


As in~\cite{Ullrich 2-electron, correlation}, it is instructive to separate $v_c^{open}(x, t)$ into an adiabatic and dynamical part,

\begin{equation}
v_c^{open}(x, t) = v_{c}^{dyn}(x, t)  + v_c^{ad}(x, t).
\end{equation}

Here,

\begin{equation}
v_c^{ad}(x, t) =  \frac{1}{4 n(x, t)}\frac{\partial^2}{\partial x^2} n(x, t) - \frac{1}{8 n(x, t)^2}\left[ \frac{\partial}{\partial x} n(x, t)\right]^2 - v_{ext}(x),
\label{adiabatic correlation}
\end{equation}

is the exact functional of ground-state DFT for one electron, or two electrons in a spin-singlet evaluated on the instantaneous OQS density~\cite{Filippi} . The dynamical part,

\begin{equation}
v_{c}^{dyn}(x, t) = - \frac{\partial}{\partial t} \alpha(x, t) - \frac{1}{2} \left[ \frac{\partial}{\partial x} \alpha(x, t)\right] ^2, 
\label{dynamical correlation}
\end{equation}

is a strictly-dynamical contribution which vanishes when the system is in thermal equilibrium. From the generalized continuity equation, eq.~\ref{pseudocontinuity}, one finds that the phase of the Kohn-Sham orbital is given by

\begin{equation}
-\frac{\partial}{\partial x} \alpha(x, t) = \frac{j(x, t)}{n(x, t)} + \frac{j_{disp}(x, t)}{n(x, t)}.
\label{orbital phase}
\end{equation}

We see that for OQS, the dynamical correlation potential $v_{c}^{dyn}(x, t)$ contains a contribution from the Hamiltonian current that is also present in usual TDDFT. In addition, there is a new contribution from the bath arising through $\textbf{j}_{disp}$. Therefore, the OQS correlation potential is a functional not only of the Hamiltonian current, but the dissipative current as well. In the following, we will consider the two limiting cases mentioned above: Pure dephasing without relaxation and relaxation without pure dephasing.

\subsection{B. Pure dephasing without relaxation}

We first consider a situation in which the OQS evolves under a Lindblad master equation which induces pure dephasing, but no relaxation~\cite{May/Kuhn, Cohen, breuer, ENAQT2}. In this case, the Lindblad operators are diagonal in a basis of eigenstates of eq.~\ref{Hamiltonian} and given by

 \begin{equation}
 L_{mn} = \delta_{mn} \sqrt{\frac{\gamma_m}{2}} |m\rangle \langle m |,
 \end{equation}
 
 where $|m\rangle$ is the mth eigenstate of the oscillator. Since the operators $L_m$ are diagonal, the populations $\rho_{mm}(t)$ remain unchanged as the system evolves. This implies that energy is conserved during the evolution, as the bath cannot drain energy away from the system. However, the coherences given by the off-diagonal density matrix elements, $\rho_{nm}(t)$, decay exponentially with a rate given by
 
 \begin{equation}
 \tau_{decoherence}^{mn} = \frac{1}{2}(\gamma_m + \gamma_n).
 \label{decoherence rate}
 \end{equation}


Pure dephasing describes a situation in which system-bath collisions are elastic, so that the bath decoheres the system without exchanging energy.




\subsection{C. Relaxation without pure dephasing}

Next, we consider the case of relaxation without pure dephasing, in which the Lindblad operators are strictly off-diagonal~\cite{May/Kuhn}. i.e.

 \begin{eqnarray}
  L_{mn} = \sqrt{\gamma_{mn}} |m\rangle \langle n|, &m \neq n& \nonumber \\
  L_{mn} = 0 ,	&m = n.&
 \label{relaxation lindblads}
 \end{eqnarray}
 
In order to ensure that the populations of the equilibrium solution obey detailed balance, one also requires that
 
 \begin{equation}
 \gamma_{mn} =  e^{\beta \omega_{mn}}  \gamma_{nm},
 \label{detailed balance}
 \end{equation}
 
 where $\beta = \frac{1}{k_B T}$ is the inverse temperature. With the Lindblad operators given by eq. \ref{relaxation lindblads}, the populations evolve according to
 
\begin{equation}
\frac{d}{dt} \rho_{nn}(t) = \sum_{m}\gamma_{nm} e^{-\beta \omega_{nm}} \rho_{mm}(t) - \rho_{nn}(t) \sum_m \gamma_{mn} e^{-\beta \omega_{mn}}.
\label{population evolution}
 \end{equation}
 
The first term in eq.~\ref{population evolution} expresses the rate at which population is transferred to $\rho_{nn}$ from all other populations $m \neq n$, while the second term expresses the rate at which population leaves $\rho_{nn}$. It can be readily verified that the right hand side of eq.~\ref{population evolution} vanishes at equilibrium where $\rho_{nn}^{eq} = \frac{e^{-\beta E_n}}{\sum_m e^{-\beta E_m}}$ and these two rates balance.
 
 The coherences evolve according to
 
 \begin{equation}
 \frac{d}{dt} \rho_{nm}(t) = \Bigg\{- \frac{1}{2} \sum_{l} \gamma_{ln}e^{-\beta \omega_{ln}} - \frac{1}{2} \sum_{l} \gamma_{lm}e^{-\beta \omega_{lm}} \Bigg\}\rho_{nm}(t).
 \label{coherence evolution}
 \end{equation}
 
This shows that even in the absence of pure dephasing, relaxation still necessarily implies decoherence. For the special case of a two-level system, one can readily see from eq's.~\ref{population evolution} and~\ref{coherence evolution} that the decoherence rate is exactly half of the relaxation rate. This is analogous to the ubiquitous phenomenological formula from NMR spectroscopy


\begin{equation}
\frac{1}{T_2} = \frac{1}{2 T_1} + \frac{1}{T_2^*},
\end{equation}

relating the time-scale for decoherence $T_2$, to that of relaxation $T_1$ and pure dephasing $T_2^*$~\cite{May/Kuhn, redfield original}. In the absence of pure dephasing, the time-scale for decoherence is twice that of relaxation, which is confirmed by eq's.~\ref{population evolution} and~\ref{coherence evolution}.
 


\section{IV. Results and analysis}

We now present and analyze the results of the inversion procedure mentioned in the previous section for obtaining the exact Kohn-Sham quantities. For all calculations, we choose the initial state of the OQS to be the pure state

\begin{equation}
|\psi(0) \rangle = \frac{1}{\sqrt{2}}(|0 \rangle + |1 \rangle),
\end{equation}

which corresponds to a density matrix with initial elements $\rho_{00}(0) = \rho_{11}(0) = \rho_{10}(0) = \rho_{01}(0) = \frac{1}{2}$ and all other entries equal to zero. The frequency of the oscillator is taken to be $\omega = 1$.

For the pure dephasing case (section III B.), we choose the parameters $\gamma_0 = \gamma_1 = 0.15$ a.u., leading to a decay of the initial coherence between $|0\rangle$ and $|1 \rangle$ at a rate of $ \tau_{decoherence}^{01} = 0.15$ a.u. as given by eq.~\ref{decoherence rate}. 

For the case of relaxation without pure dephasing (section III C.), we choose the population transfer rate from $|1\rangle$ to $|0 \rangle$ to be $\gamma_{01} = 0.3$ a.u. and the inverse temperature to be $\beta = 1$. Using eq.~\ref{detailed balance}, this gives $\gamma_{10} = \frac{\gamma_{01}}{e} \approx 0.11$ a.u. and $\gamma_{20} =  \frac{\gamma_{01}}{e^2} \approx 0.04$ a.u. At this relatively low temperature, the transfer rates to excited states higher than $|2 \rangle$ are sufficiently small that they can be neglected. Using eq.~\ref{coherence evolution}, the decay of the initial coherence between $|0\rangle$ and $|1 \rangle$ is found to be $\tau_{decoherence}^{01} = \frac{\gamma_{01}}{2}(1+ \frac{1}{e}) \approx 2.1$ a.u. These parameters all correspond to underdamped motion in which the decay to equilibrium occurs on a much longer time-scale than the oscillation period of the system. This condition is implicit in the assumption of weak system-bath coupling.

\subsection*{A. The density, current and dissipative current}

\begin{figure}[htbp]
\begin{center}
\includegraphics[width = 75mm, height = 110mm]{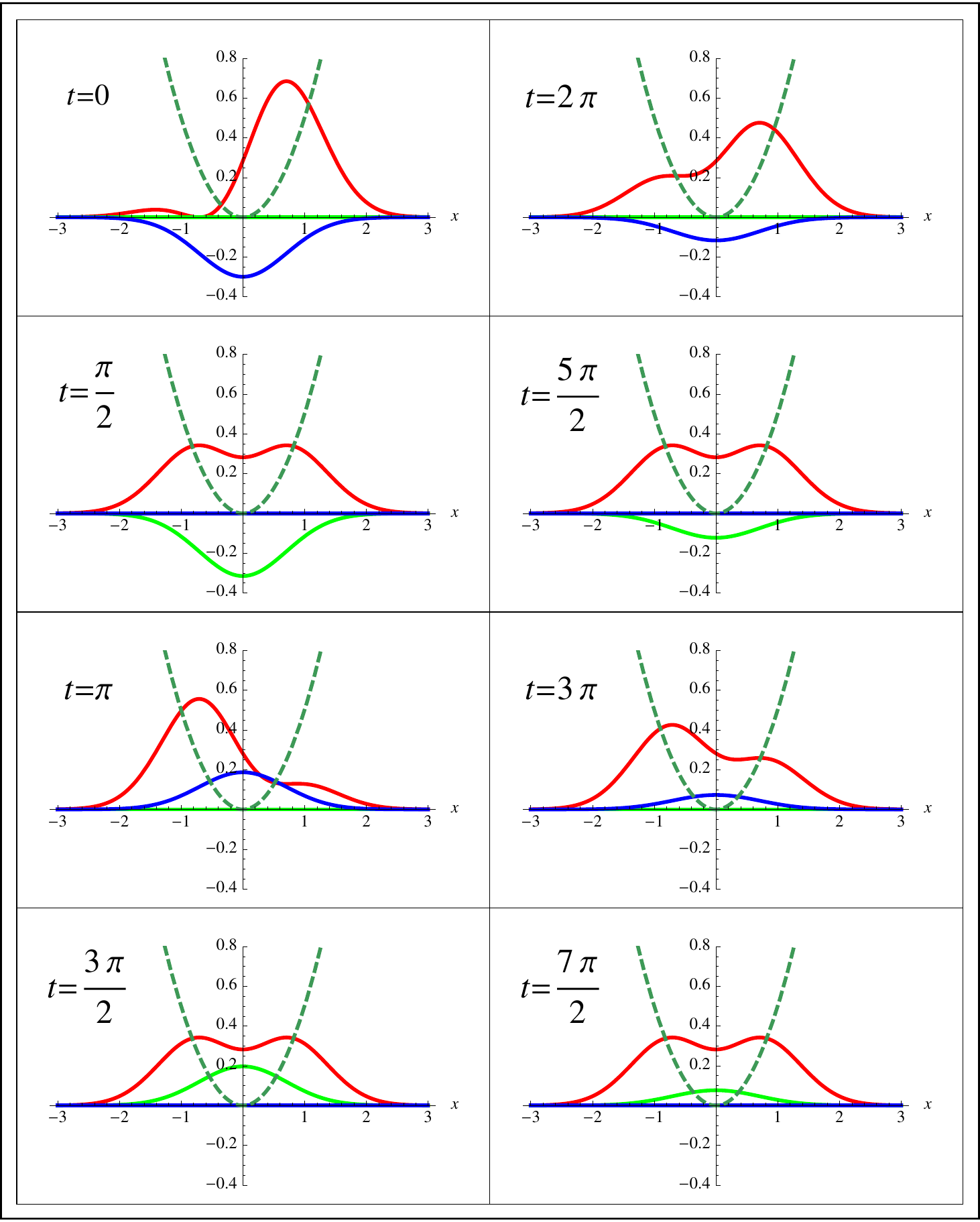}
\caption{Time evolution of the density $n(x,t)$ (red-solid), the Hamiltonian current $j(x,t)$ (green-dot-dashed) and the dissipative current $j_{disp}(x,t)$ (blue-dotted), for the pure dephasing master equation. $j_{disp}(x,t)$ has been scaled by a factor of 5 to make it more visible in the figure. $v_{ext}(x) = \frac{1}{2} \omega^2 x^2$ (green-dashed) is shown for reference as well.}
\label{puredephasingdensities}
\end{center}
\end{figure}

\begin{figure}[htbp]
\begin{center}
\includegraphics[width = 75mm, height = 110mm]{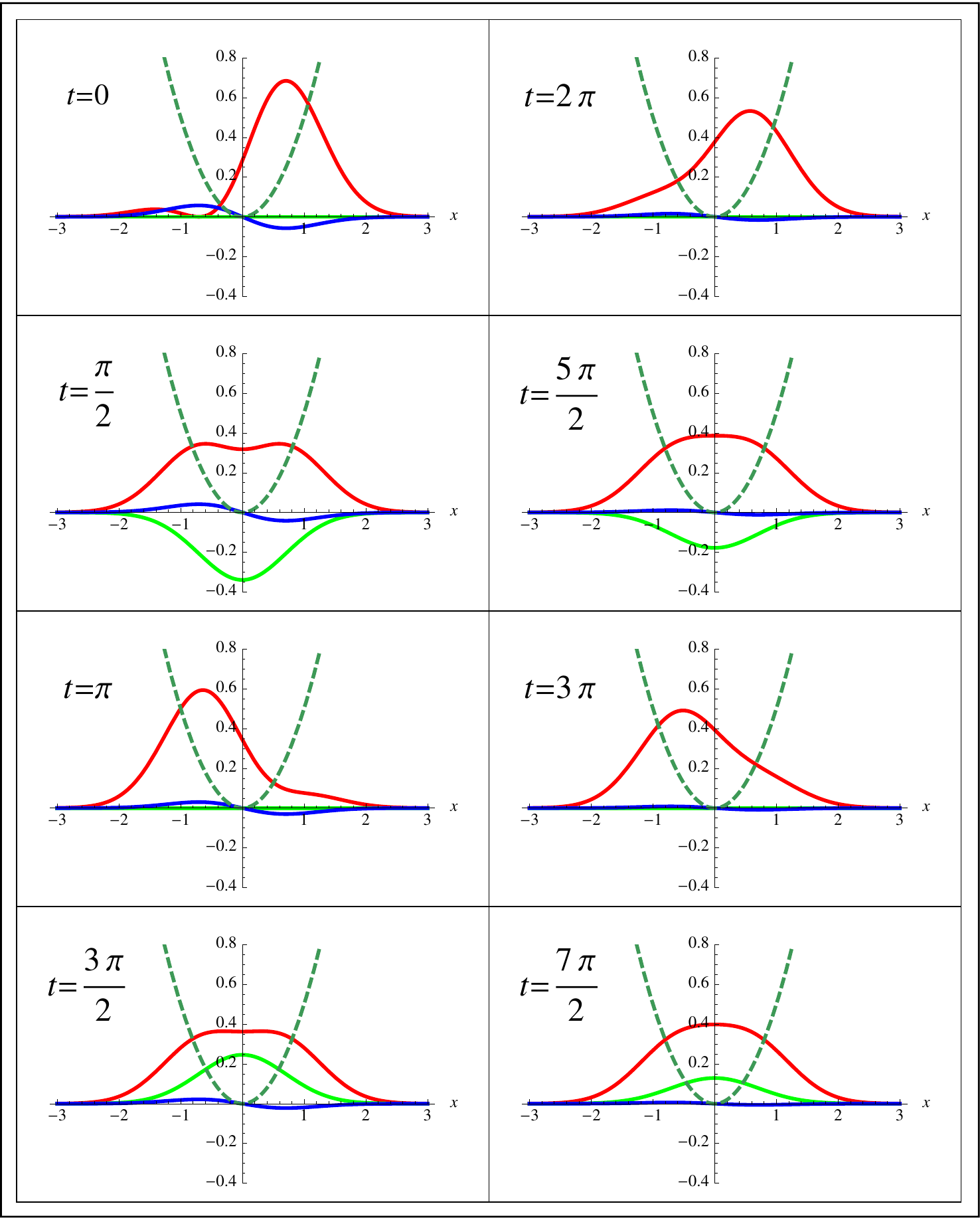}
\caption{Time evolution of the density $n(x,t)$ (red-solid), the Hamiltonian current $j(x,t)$ (green-dot-dashed) and the dissipative current $j_{disp}(x,t)$ (blue-dotted), for the master equation including relaxation with no pure dephasing. $j_{disp}(x,t)$ has been scaled by a factor of 5 to make it more visible in the figure. $v_{ext}(x) = \frac{1}{2} \omega^2 x^2$ (green-dashed) is shown for reference as well.}
\label{relaxationdensities}
\end{center}
\end{figure}

In Figures~\ref{puredephasingdensities} and~\ref{relaxationdensities}, we plot the density, current density and dissipative current  density as a function of time for the pure dephasing and relaxation without pure dephasing master equations respectively. Since we have chosen parameters corresponding to weak system-bath coupling, the dissipative current $j_{disp}(x, t)$ is significantly smaller than the Hamiltonian current $j(x, t)$, by approximately a factor of $ \tau_{decoherence}^{01}$ for the pure dephasing case. In the figures, we have multiplied $j_{disp}(x, t)$ by a factor of 5 to make it more visible.

In Figure~\ref{puredephasingdensities}, we see that the coherences decay exponentially in time as the density oscillates in the harmonic well, while the populations remain unchanged, as expected for pure dephasing. In the absence of coupling to the bath, time translational invariance would imply that the snapshots in the right hand column would be the same as those on the left, which occur at a time $t = 2\pi$ earlier. The loss of coherence manifests itself in the decay of the currents and changing density profile as the system evolves toward the fully mixed state $\rho_{00} = \rho_{11} =\frac{1}{2}$, $\rho_{10} = \rho_{01} = 0$ at equilibrium, which is an incoherent sum of the ground and first excited states. One can also see that the dissipative current is proportional to the spatial integral of the real part of the coherence, and lags the hamiltonian current by a phase of $\frac{\pi}{2}$ which is proportional the spatial integral of the imaginary part. Only the real part of the coherence contributes to the density evolution.

The contribution to the density from the coherences is antisymmetric about the origin, while the contribution from the populations is symmetric. When these two contributions are superposed, the density acquires an asymmetric profile, which can be seen at integer multiples of $\pi$ where the coherent contribution is a maximum. At half integer multiples of $\pi$, the contribution from the real part of the coherences instantaneously vanishes and the density becomes symmetric. In contrast, the currents are always perfectly symmetric about the origin since they are proportional only to the spatial integral of the antisymmetric coherences.

In Figure~\ref{relaxationdensities}, we see that in addition to decoherence there is population transfer, since we have included energy relaxation in the master equation. By $t=\frac{7 \pi}{2}$, the system has already nearly approached the equilibrium state

\begin{equation}
\hat{\rho}_S(t = \infty) = \frac{1}{1+ e^{-\beta \omega}} |0 \rangle \langle 0| + \frac{e^{-\beta \omega}}{1+ e^{-\beta \omega}} |1 \rangle \langle1|,
\end{equation}

which for the parameters we have chosen corresponds to $\rho_{00}(t = \infty) \approx 0.73$ and $\rho_{11}(t = \infty) \approx 0.27$. Since the equilibrium density is dominated by the ground-state, it is nearly gaussian, but slightly flattened due to a mixing in of the first excited state. The dissipative current has a similar profile to that seen in~\cite{gebauer prl}, which results from the fact that population transfer generates no Hamitonian current, but does give rise to a change in the density which must be compensated by $j_{disp}(x, t)$.

\subsection*{B. The exact Kohn-Sham potential, Kohn-Sham orbitals and density matrices}

\begin{figure}[htbp]
\begin{center}
\includegraphics[width = 75mm, height = 110mm]{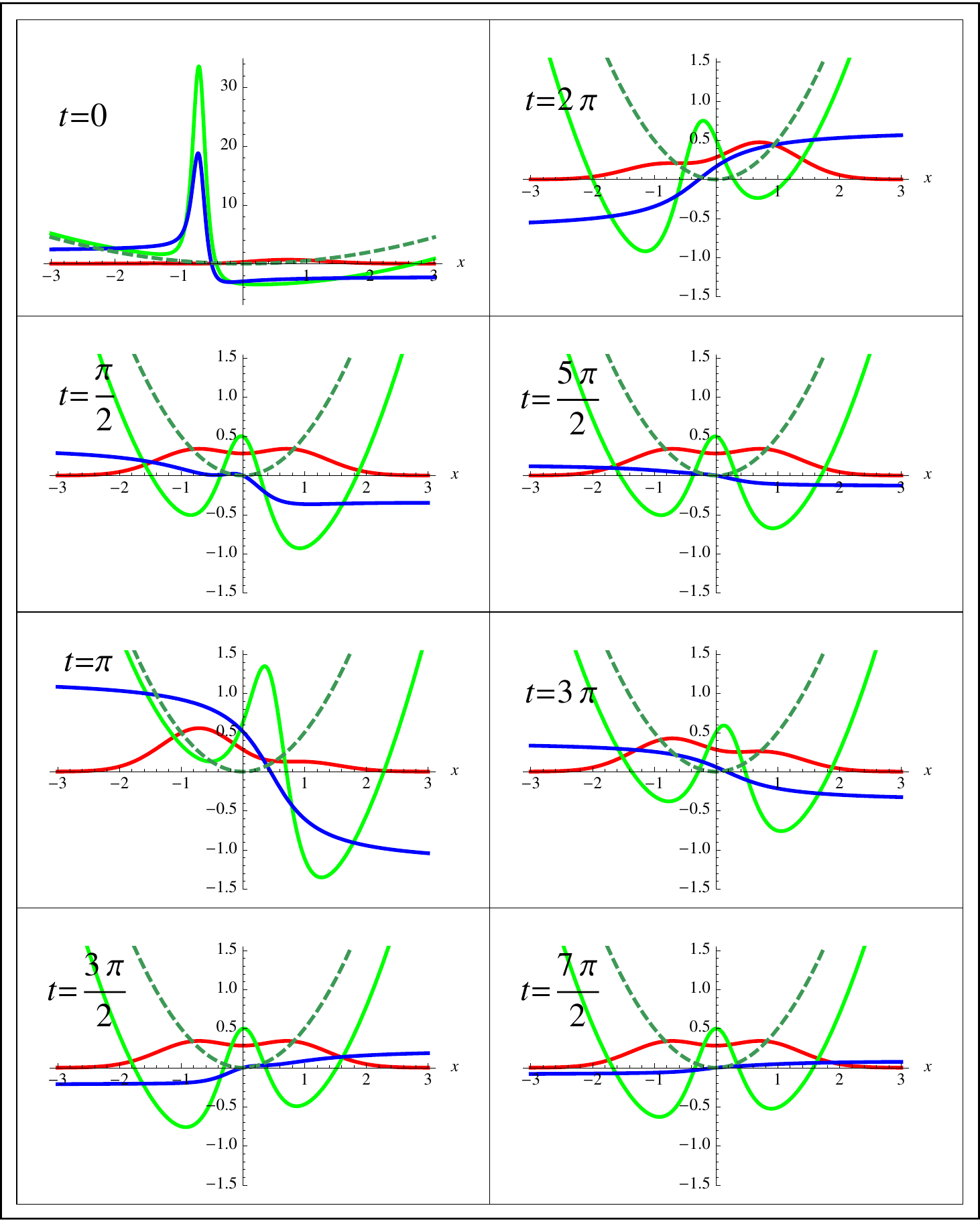}
\caption{Time evolution of the density $n(x,t)$ (red-solid), the full Kohn-Sham potential $v_{ks}(x,t) = v_c^{open}(x, t) + v_{ext}(x)$ (green-dot-dashed) and the dynamical correlation potential $v_{c}^{dyn}(x, t)$ (blue-dotted) for the pure dephasing master equation. Note the change of scale for the $t=0$ frame. $v_{ext}(x) = \frac{1}{2} \omega^2 x^2$ (green-dashed) is shown for reference as well.} 
\label{puredephasingpotentials}
\end{center}
\end{figure}

\begin{figure}[htbp]
\begin{center}
\includegraphics[width = 75mm, height = 110mm]{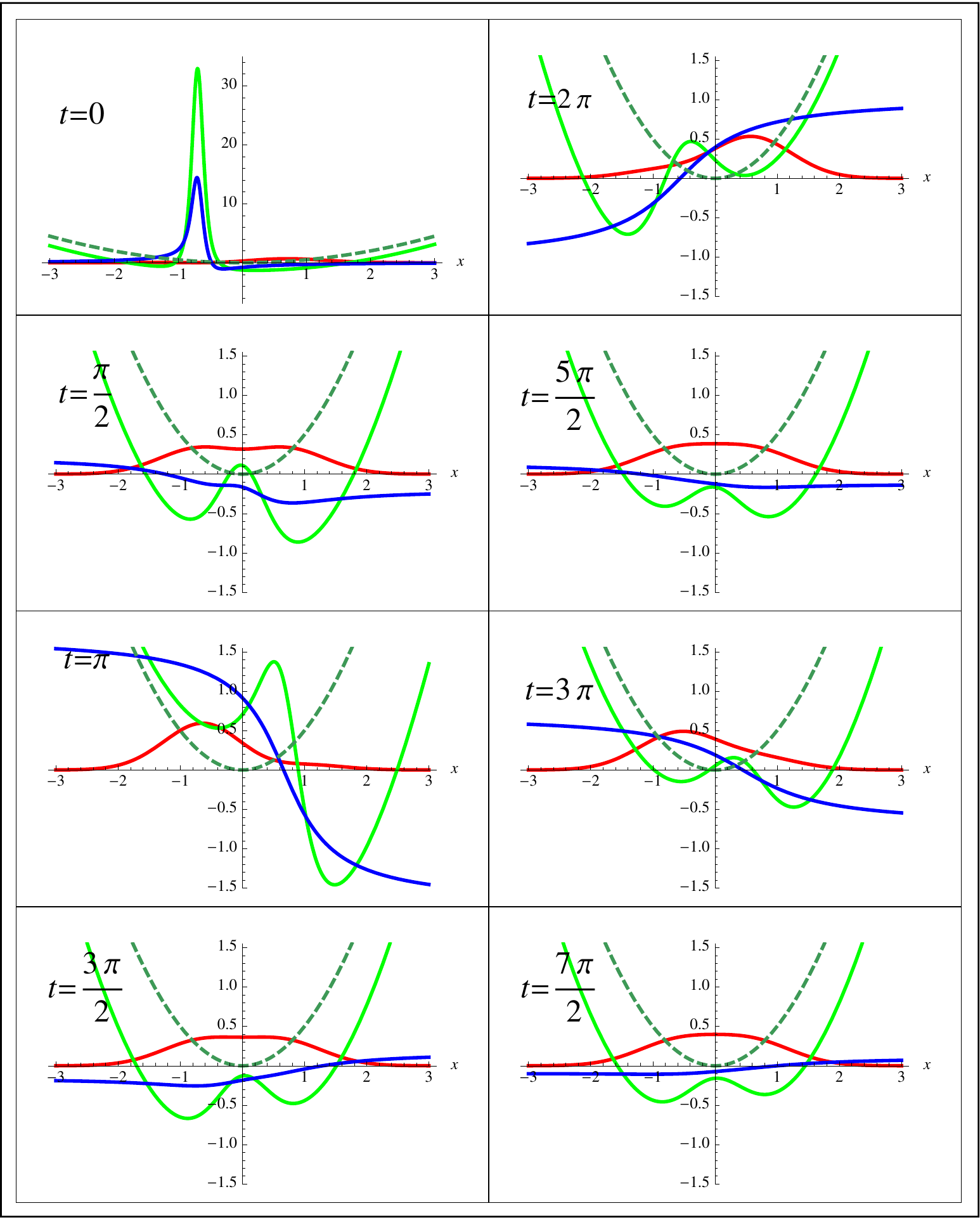}
\caption{Time evolution of the density $n(x,t)$ (red-solid), the full Kohn-Sham potential $v_{ks}(x,t) = v_c^{open}(x, t) + v_{ext}(x)$ (green-dot-dashed) and the dynamical correlation potential $v_{c}^{dyn}(x, t)$ (blue-dotted) for the master equation including relaxation with no pure dephasing. Note the change of scale for the $t=0$ frame. $v_{ext}(x) = \frac{1}{2} \omega^2 x^2$ (green-dashed) is shown for reference as well.}
\label{relaxationpotentials}
\end{center}
\end{figure}

\begin{figure*}[htbp]
\begin{center}
\includegraphics[width = 185mm, height = 70mm]{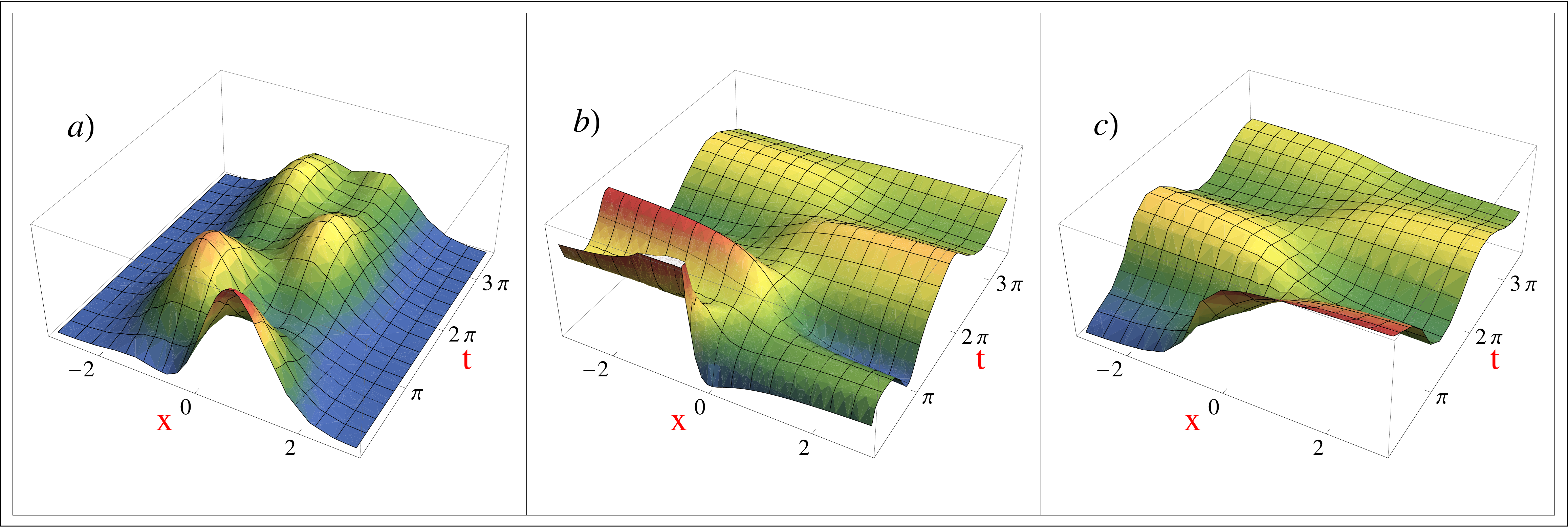}
\caption{a) The density n(x,t),  b)  the exact dynamical correlation potential $v_{c}^{dyn}(x, t)$ and c) the Markovian bath functional (MBF), $v_{c}^{dyn, MBF}(x, t)$ for the pure dephasing master equation during the entire duration of the simulation shown as a function of space and time.}
\label{3-dplot}
\end{center}
\end{figure*}

In Figures~\ref{puredephasingpotentials} and~\ref{relaxationpotentials}, we show the full Kohn-Sham potential and its dynamical contribution (eq.~\ref{dynamical correlation}) for the density evolution presented in Figures~\ref{puredephasingdensities} and~\ref{relaxationdensities} respectively. In Figure~\ref{3-dplot}, we plot the density and dynamical potential $v_{c}^{dyn}(x, t)$ as a function of space and time for the pure dephasing case as well. Examining Figures~\ref{puredephasingpotentials} and~\ref{3-dplot}, we see that for pure dephasing, the dynamical potential becomes extremely repulsive in regions where the contribution to the density arising from the coherences in the OQS density matrix is large. For instance, at $t = \pi$, the contribution to the density from the coherences is positive for $x<0$ and negative for $x>0$. Likewise, $v_{c}^{dyn}(x, t)$ is positive for $x<0$ and negative for $x>0$, which corresponds to a field which will tend to suppress the coherent part of the density. For $t = 2\pi$, the sign of the coherent part of the density is reversed and  $v_{c}^{dyn}(x, t)$ is as well, again corresponding to a potential which redistributes the density so as to suppress its coherent contribution. This behavior gives us some insight into how the Kohn-Sham system is able to use a unitary evolution to mimic the effect of the bath. In the Kohn-Sham system, decoherence is effectively converted into a control problem, in which the local time-dependent field $v_{c}^{dyn}(x, t)$ drives the density in the same way that collisions with the bath would in the true open system. This is a complicated task, since not only must $v_{c}^{dyn}(x, t)$ drive down the coherences, but it must do so without affecting the contribution to the density from the populations. The situation is somewhat different in Figure~\ref{relaxationpotentials}, since now the potential not only suppresses the coherent density evolution, but it also transfers population from the first excited state to the groundstate to reach the thermal equilibrium density distribution. 

At $t=0$, the Kohn-Sham potentials appear to be extremely large, however this occurs in a region of space where the density vanishes and has little effect on the dynamics. At long times, one sees that the dynamical potential decays, while in the pure dephasing case, the full Kohn-Sham potential acquires a double well structure. This corresponds to the adiabatic Kohn-Sham potential (eq.~\ref{adiabatic correlation}) evaluated on the fully mixed, equilibrium-state density. In figure~\ref{relaxationpotentials}, the double well structure is less pronounced, since the equilibrium state is dominated by the groundstate for which the adiabatic potential reduces to $v_{ext}$ (up to a constant), which is parabolic.

\begin{figure}[htbp]
\begin{center}
\includegraphics[width = 75mm, height = 110mm]{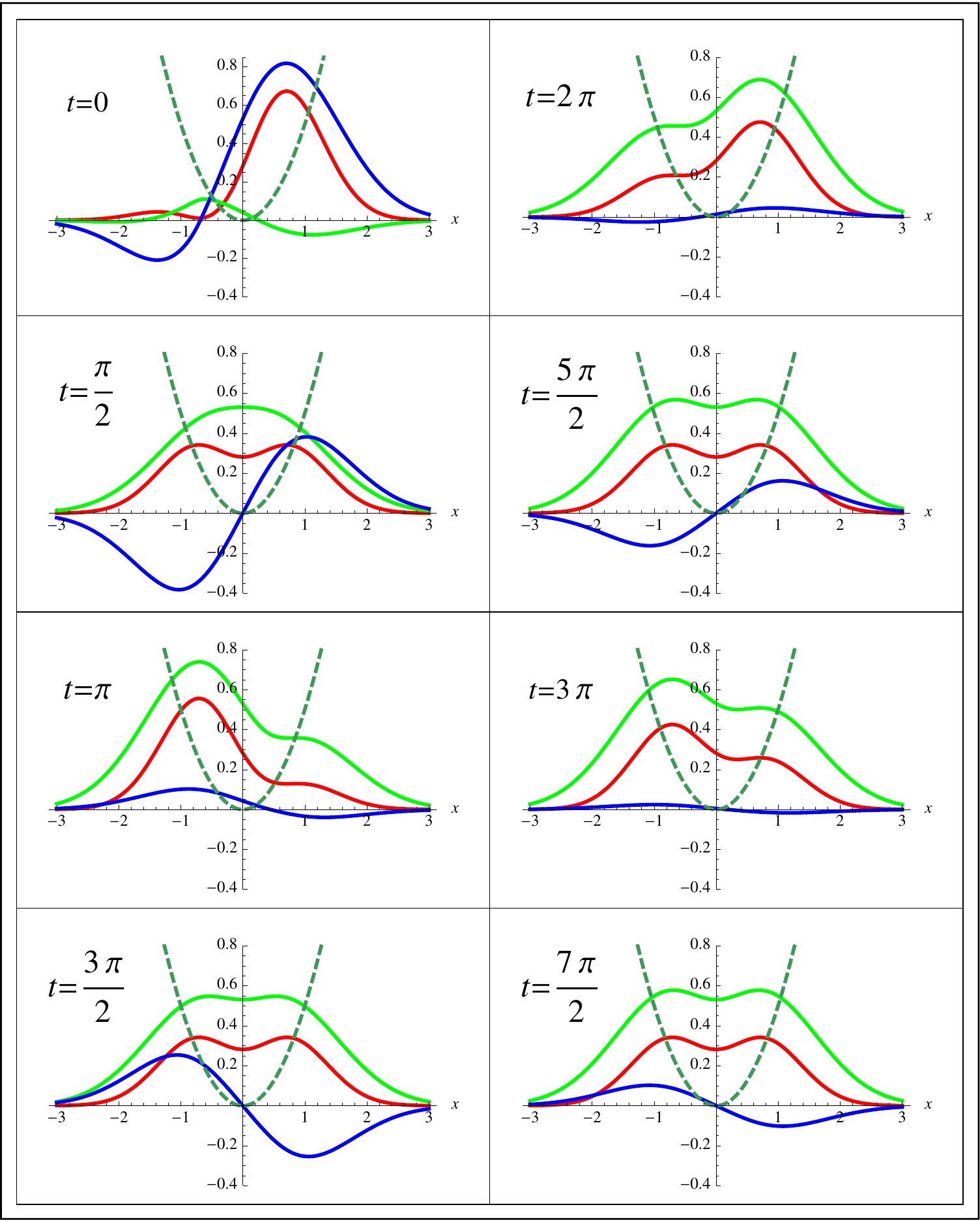}
\caption{Time evolution of the density $n(x,t)$ (red-solid), the real part of the Kohn-Sham orbital $\Re e \phi(x, t)$ (green-dot-dashed) and the imaginary part of the Kohn-Sham orbital $\Im m \phi(x, t)$ (blue-dotted), for the pure dephasing master equation. $v_{ext}(x) = \frac{1}{2} \omega^2 x^2$ (green-dashed) is shown for reference as well.}
\label{puredephasingorbitals}
\end{center}
\end{figure}

\begin{figure}[htbp]
\begin{center}
\includegraphics[width = 75mm, height = 110mm]{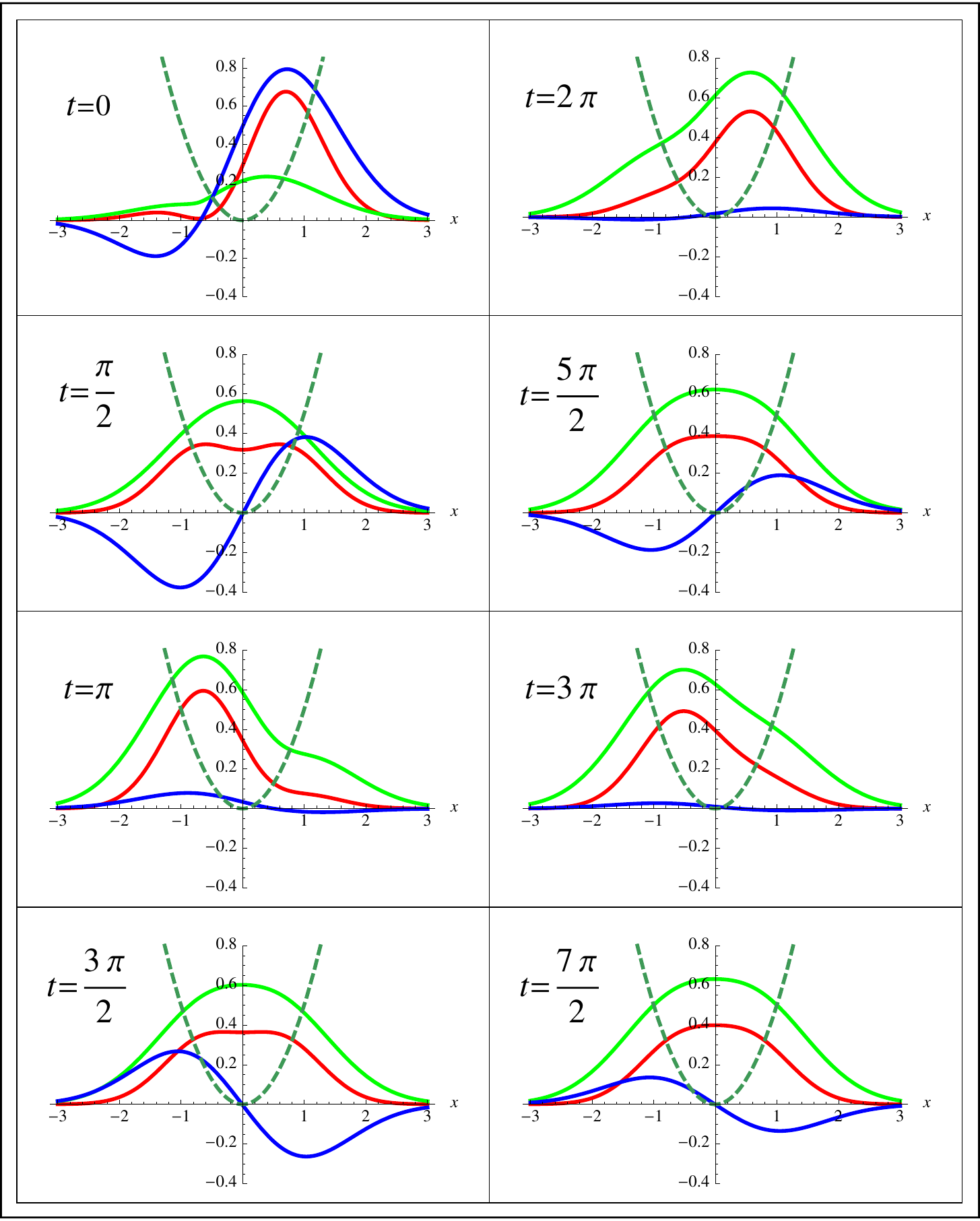}
\caption{Time evolution of the density $n(x,t)$ (red-solid), the real part of the Kohn-Sham orbital $\Re e \phi(x, t)$ (green-dot-dashed) and the imaginary part of the Kohn-Sham orbital $\Im m \phi(x, t)$ (blue-dotted), for the master equation including relaxation with no pure dephasing.  $v_{ext}(x) = \frac{1}{2} \omega^2 x^2$ (green-dashed) is shown for reference as well.}
\label{relaxationorbitals}
\end{center}
\end{figure}

In Figures~\ref{puredephasingorbitals} and~\ref{relaxationorbitals}, the Kohn-Sham orbital corresponding to the two different master equations is shown. The imaginary part of the orbital arises from the Hamitonian current as in usual TDDFT, but also has a contribution from the dissipative current. At long times, the currents decay and so the imaginary part vanishes while the real part is simply given by $\sqrt{n(x,t)}$. In the long-time limit, this orbital must reduce to an eigenstate of the adiabatic Kohn-Sham potential evaluated on the equilibrium OQS density. In this way, the Kohn-Sham system reproduces the correct equilibrium density of a mixed-state density matrix using a pure-state wavefunction. In principle, this wavefunction need not be the groundstate, but it is in the cases studied here.

\begin{figure}[htbp]
\begin{center}
\includegraphics[width = 75mm, height = 110mm]{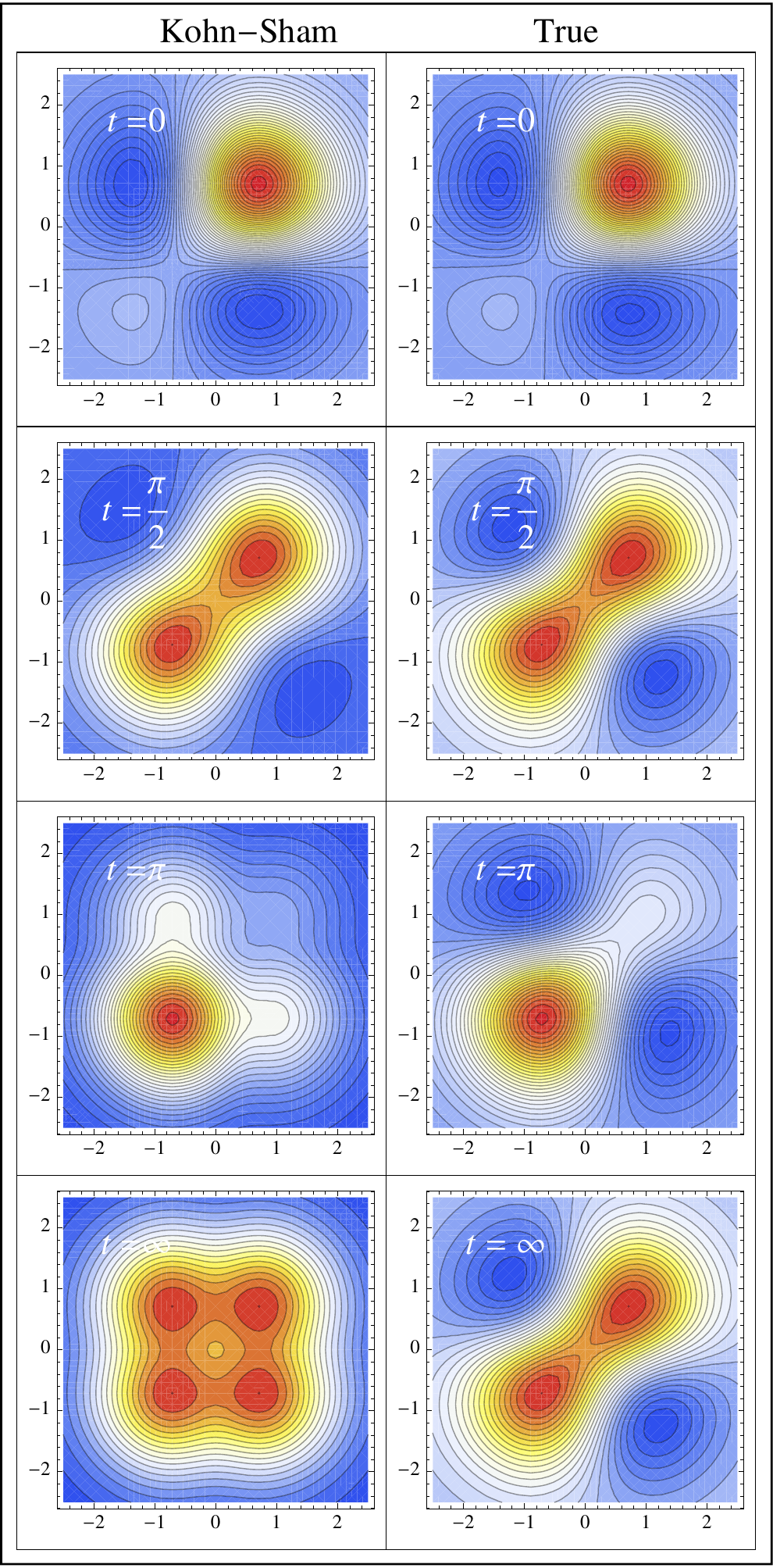}
\caption{Time evolution of the real part of the Kohn-Sham density matrix $\Re e \rho_{ks}(x, x', t)$ (left column) and true OQS density matrix $\Re e \rho_{open}(x, x', t)$ (right column) for the pure dephasing master equation.}
\label{puredephasingmatrixwh}
\end{center}
\end{figure}

\begin{figure}[htbp]
\begin{center}
\includegraphics[width = 75mm, height = 110mm]{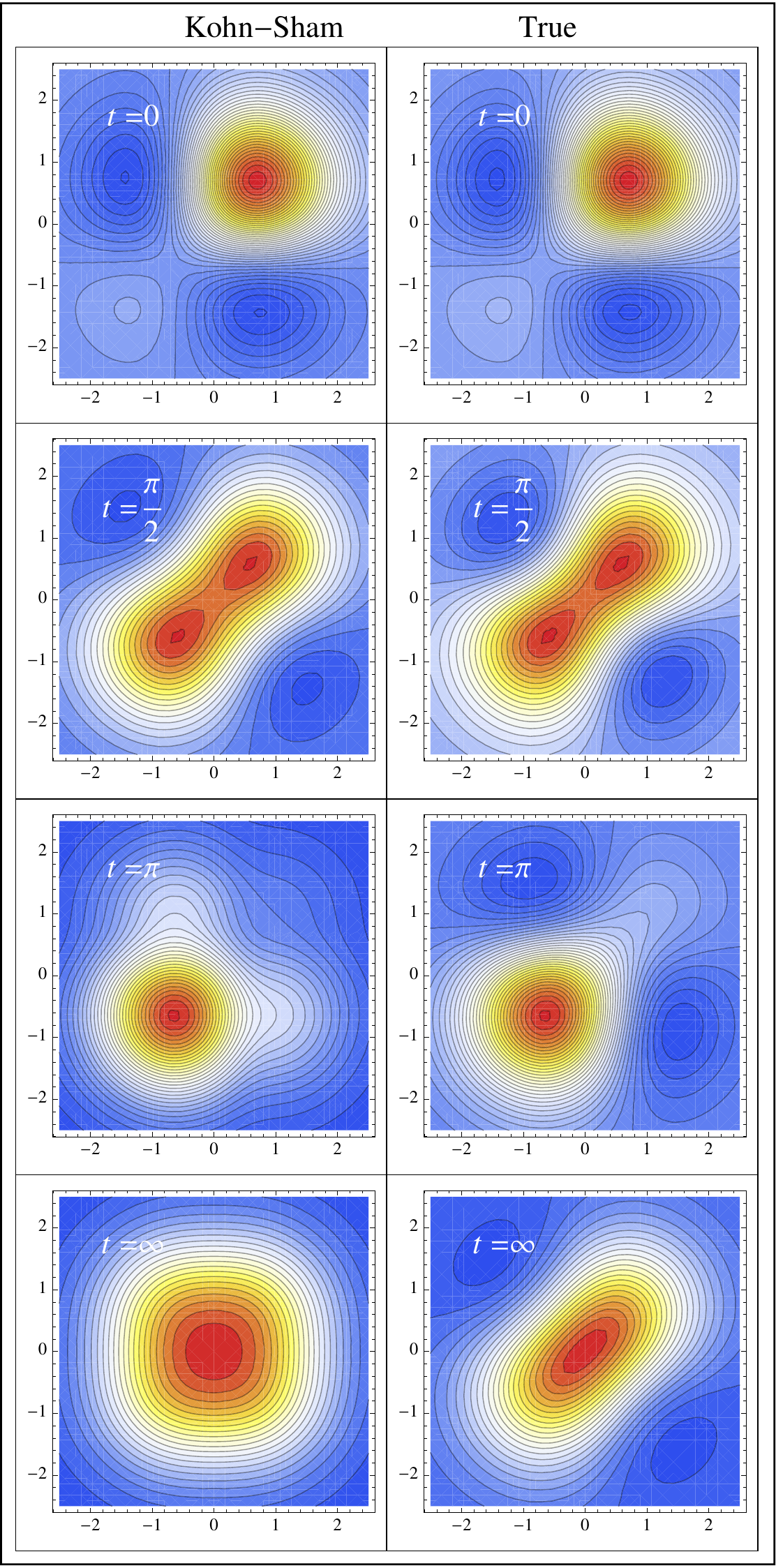}
\caption{Time evolution of the real part of the Kohn-Sham density matrix $\Re e \rho_{ks}(x, x', t)$ (left column) and true OQS density matrix $\Re e \rho_{open}(x, x', t)$ (right column) for the master equation including relaxation with no pure dephasing.}
\label{relaxationmatrixwh}
\end{center}
\end{figure}

Lastly, we show the real part of the real-space density matrices for both the Kohn-Sham system and the true OQS in Figures~\ref{puredephasingmatrixwh} and~\ref{relaxationmatrixwh}. In the Kohn-Sham case, this is simply the pure-state density matrix

\begin{equation}
\rho_{ks}(x, x', t) = \phi^*(x,t) \phi(x't),
\end{equation}

while in the OQS case it is the position representation of the solution to eq. \ref{HO lindblad},

\begin{equation}
\rho_{open}(x, x', t) = \langle x| \hat{\rho}_S(t)| x' \rangle.
\end{equation}

Since we start in a pure-state at $t=0$, the Kohn-Sham density matrix coincides with that of the true OQS. As the system evolves, the two begin to differ as the true density matrix looses purity while that of the Kohn-Sham system does not. However, it can be seen that at all times $\rho_{ks}(x, x, t) = \rho_{open}(x, x, t) =n(x, t)$ and the Kohn-Sham system reproduces the true OQS density on its diagonal.

In both Figure~\ref{puredephasingmatrixwh} and Figure~\ref{relaxationmatrixwh}, we see that the Kohn-Sham density matrix has large off-diagonal support which is not present in the true density matrix. This is consistent with the fact that the Kohn-Sham density matrix must remain idempotent. i.e.

\begin{equation}
\int dx \int dx' |\rho_{ks}(x, x', t)|^2 = 1,
\label{idempotency}
\end{equation}

while for the true system,

\begin{equation}
\int dx \int dx' |\rho_{open}(x, x', t)|^2 < 1.
\end{equation}

The Kohn-Sham density matrix is able to satisfy eq. \ref{idempotency} by increasing the off-diagonal regions and thus compensating for the lost volume in the true density matrix without affecting the density.

\subsection*{C. The Markovian bath functional}

Our analysis of the exact OQS-TDDFT functional in section IV B also sheds some light on the Markovian bath functional (MBF) that we first presented in~\cite{prl}. In the MBF, one approximates the dynamical correlation potential as

\begin{equation}
v_{c}^{dyn}(x, t) \approx v_{c}^{dyn, MBF}(x, t) \equiv \lambda \int_{-\infty}^{x} \frac{j(x',t)}{n(x',t)} dx',
\end{equation}

where $\lambda$ is a fitted parameter. In this approximation, one neglects the contribution to $v_{c}^{dyn}(x, t)$ arising from the second term in eq.~\ref{dynamical correlation} involving gradients of the phase \textit{and} one also neglects the contribution from the dissipative current. The first approximation is valid when the phases are spatially slowly varying, while the second approximation is valid if the coupling to the bath is weak. Lastly, one assumes

\begin{equation}
\frac{d}{dt} \alpha(x,t) \propto -\lambda \alpha(x,t).
\end{equation}

i.e. the phase of the orbital is exponentially damped, which is valid in the Markovian limit. The MBF is a simple yet practical functional as it only involves knowledge of the Hamiltonian current and density at each instant in time. It can readily be implemented in real-time codes with existing propagation schemes~\cite{neuhauser}. Numerical simulations  of dissipation of excited states of a model Helium atom show promising results, notwithstanding the simplicity of the MBF~\cite{prl}.

In Figure~\ref{3-dplot}, we show a comparison of the MBF  $v_{c}^{dyn, MBF}(x, t)$ and the exact functional $v_{c}^{dyn}(x, t)$ for the pure dephasing master equation of the model system presented in the previous sections. The fitted decay parameter is taken to be the decoherence rate in the master equation. i.e. $\lambda =   \tau_{decoherence}^{01} = 0.15$. This is natural since it is the only decoherence timescale in the problem. The MBF is seen to reproduce the long-time behavior of the exact dissipative potential reasonably well, however it deviates significantly for short times. This arises from the fact that the Markov approximation only provides an adequate description of the dynamics for times longer than the decoherence time. At much shorter times it is not in general valid. In Figure~\ref{3-dplot}, we see that this is roughly the timescale for the MBF to become accurate. Also, because the MBF neglects the dissipative current which is out of phase with the Hamiltonian current, there is a slight phase shift in the MBF with respect to the exact dissipative functional which contains both current contributions.

\section{V. Conclusion and Outlook}

In summary, we have explored the behaviour of the exchange-correlation potential for dissipative open quantum systems using an exactly-solvable one-electron system. The two limiting cases (pure dephasing and relaxation) provided insights into the time-dependence and form of the dissipative potential that will need to be described in OQS-TDDFT functionals. We explored the behavior of the closed auxiliary Kohn-Sham system, as the system evolves under decoherence and relaxation. This is valuable information for the development of many-body, realistic bath functionals for OQS-TDDFT. 

 We have seen that the Kohn-Sham system mimics the effect of a bath by unitarily evolving with a dynamical potential, which depends on both the Hamitonian and dissipative currents. For a pure dephasing master equation,  this potential tends to suppress the coherent part of the density evolution in a delicate way so that energy is still conserved and populations remain unchanged. When relaxation is also present, the potential drains energy away from the system and reaches an equilibrium distribution which satisfies detailed balance.

When dealing with a many-electron system, one expects that the functionals should retain some of the same features as in the one-electron case studied here, however, there are some fundamental differences. Firstly, one does not know the many-electron eigenstates \textit{a priori} and so it is generally not possible to explicitly write down Lindblad operators in terms of these eigenstates as was done in eq.~\ref{HO lindblad}. Instead, one should start from a master equation written in terms of effective single-particle eigenstates, such as a basis of orbitals from an equilibrium-state Kohn-Sham-Mermin calculation~\cite{kohn-sham, burke, tempel}. 

Secondly, the simple form of the dynamical functional expressed in eq.~\ref{dynamical correlation} no longer holds. In the simulation we have considered here, there are only two relevant levels involved in the dynamics. In general, for a many-level system the bath will induce different relaxation and dephasing rates for different eigenfrequencies of the system, depending on the bath spectral density~\cite{May/Kuhn, van Kampen, breuer}. Rather than simply coupling to the currents in a semi-time-local manner as in eq.~\ref{dynamical correlation}, the exact functional will need to damp different Fourier components of the current at different rates. In the time-domian, this corresponds to a complicated memory-dependence on both the Hamiltonian and dissipative currents at earlier times. The form of this memory-dependence should clearly depend on the bath spectral density, but its exact structure will need to be investigated in future work. 

In some respects, OQS-TDDFT functionals are similar to existing current-dependent functionals in TDCDFT, where frequency-dependent dissipation arises due to coupling to currents at earlier times via a stress tensor~\cite{vignale springer, Vignale-Kohn, VUC, Vignale-Dgosta, harshani}. However, the physical origin of the dissipation is very different in the two theories. In TDCDFT, the frequency dependence of the stress tensor depends on viscoelastic coefficients derived from the uniform electron gas. In contrast, the OQS-TDDFT functional depends on the spectral density of a bosonic bath such as phonons or photons.

We acknowledge NSF award PHY-0835713 for financial support. A.A.G. acknowledges support from the Camille and Henry Dreyfus Teacher-Scholar award, and the Sloan Research Fellowship. We would also like to thank J. Yuen-Zhou and John Parkhill for helpful discussions.

\end{document}